\documentclass[prl,twocolumn,amsmath,amssymb,superscriptaddress]{revtex4}

\usepackage{graphicx}
\usepackage{dcolumn}
\usepackage{amsmath}
\usepackage{amssymb}

\newcommand{\be}{\begin{equation}}
\newcommand{\ee}{\end{equation}}
\newcommand{\bea}{\begin{eqnarray}}
\newcommand{\eea}{\end{eqnarray}}
\newcommand{\p}{\partial}

\begin{document}

\title{Unstable fingering patterns of Hele-Shaw flows as a dispersionless limit of
the KdV hierarchy}

\author{Razvan Teodorescu}
\affiliation{Physics Department, Columbia University, 538 W.
120th St., New York, NY 10027}
\author{Paul Wiegmann}
\affiliation{James Frank Institute, Enrico Fermi Institute of the
University of Chicago,   5640 S. Ellis Ave.
Chicago IL 60637}
\altaffiliation[Also at ]{Landau Institute of Theoretical Physics.}
\author{Anton Zabrodin }
\affiliation{Institute of Biochemical Physics, Kosygina str. 4,
117334 Moscow, Russia}
\altaffiliation[Also at ]{ITEP, Moscow, Russia.}

\date{\today}

\begin{abstract}
We show that unstable fingering patterns of two dimensional flows of
viscous  fluids with open boundary  are  described by a 
dispersionless limit of the KdV hierarchy.
In this framework, the
fingering instability is linked to a known instability leading to   
regularized shock
solutions for nonlinear waves, in dispersive  media. The  integrable
structure of the flow  suggests a dispersive regularization of the finite-time
singularities.
\end{abstract}

\maketitle

\paragraph*{1. Introduction. }The Hele-Shaw cell is a narrow space between two
parallel planes -- a device used to study the 2D dynamics of a fluid with an open
boundary.  In a common set-up, air (regarded as a non-viscous fluid) occupies a bounded  domain of  the cell,
 otherwise filled by a viscous incompressible fluid.  When more air is
injected through a well, the free boundary evolves in a complicated and
unstable manner.  In finite time, an arbitrary smooth initial boundary develops
a pattern of  branched fingers
. This 
mechanism is 
rather general and has been identified in numerous 
growth problems 
\cite{Hele-Shaw}.

Starting from early works \cite{Kochina}, it became clear that the
problem of fingering instability in the Hele-Shaw cell
is linked  to profound aspects
of  the analytical functions theory and 2D conformal maps.
In recent years, the problem has reached other domains of theoretical physics.
In particular, a connection to fractal self-similar patterns
 of stochastic growth, like
DLA, has been recognized \cite{DLA}.

The Hele-Shaw problem has also emerged
in the context of
 electronic physics in low dimensions.
For 1D electrons, the Fermi sphere
is deformed by external perturbations
according to the Hele-Shaw mechanism \cite{Wiegmann}.
The same is true for deformations
of an  electronic droplet in a quantizing magnetic field \cite{ABWZ}.
Similarly, the Hele-Shaw problem is  connected to statistical
ensembles of normal
or complex random matrices, where a non-Gaussian potential
   deforms the support of eigenvalues  \cite{TBAZW}.

In this paper we emphasize yet another  connection, which may be viewed as
 a formal ground for  appearance of this phenomenon in different physical
situations.   We show that the  fingering instability
is linked to the integrable
Korteweg-de Vries (KdV) hierarchy of differential equations. More precisely,
 we show that developed fingers, in the
absence of surface tension, are described by the dispersionless KdV
hierarchy.
The Hele-Shaw
fingers correspond to the same solution of the KdV
hierarchy as that exploited in  2D-gravity \cite{BK}, and in studies of critical
points  of Hermitian random matrix ensembles \cite{DiFrancesco}.

The integrable structure of the Hele-Shaw problem has been
observed in our earlier
  papers \cite{MWZ}.  There, we have shown that the problem is equivalent
  to the dispersionless limit of the Toda integrable hierarchy.
This is an exact relation.
Here we concentrate  on  the critical  (turbulent) regime of
  the flow, where fingers
are already well developed, and 
close to a cusp-like singularity.  
 In this case,  one may
concentrate  only on the vicinity of
the tip of a single finger, neglecting the rest of the boundary.
The KdV integrable structure
emerges in this regime. It can be obtained from the Toda hierarchy as a reduction, but can also
 be derived directly from hydrodynamics of the critical regime. We choose the latter.


In this letter, the intuitive physics
of interface dynamics   gives a clear
geometrical  interpretation to algebraic objects used 
in dispersionless soliton equations \cite{Takasake}.

\paragraph*{2. Darcy's law.}  Consider a 2D domain
(a ``bubble") occupied by ``air" regarded as an
incompressible fluid with low viscosity.
  The rest of the cell is occupied by
another incompressible fluid but with high viscosity. Air is injected
into the cell through the origin,  at a constant rate,
while the viscous fluid is evacuated from ``infinity" (edges
of the cell). The area of the bubble  is proportional to time $t$.
We normalize it to be $\pi t$. The Navier-Stokes equation
adapted to the 2D geometry
gives a  simple rule for  the dynamics of the moving boundary $\gamma$:
velocity
in the viscous fluid
(and of the boundary) is proportional to the gradient of pressure,
\begin{equation}\label{1}
\vec{v} = -\vec\nabla p.
\end{equation}
Pressure is harmonic for incompressible fluids.  Inside the bubble, it
is constant (set to $p=0$), due to low viscosity.
In the absence of surface tension,
it is continuous across the boundary, and hence
 a solution of  the
problem
\begin{equation}\label{2}
\Delta p=0,\quad p \bigl |_{\infty} \to -\log |z|, \quad
p \bigl |_{\gamma }=0.
\end{equation}
The origin of the fingering instability is intuitively clear --
high curvature portions of the
  boundary move with a higher velocity than the rest and get even more curved.

We now introduce a minimal set of  objects of
  potential theory. Consider a   holomorphic function
$\phi(z)=\xi(x,y)+ip(x,y)$,
whose imaginary part is pressure. The function $\phi(z)$ is a
univalent conformal  map of the exterior of the bubble to the cylinder
$Im\phi > 0$, and its derivative $\p\phi(z)$ taken on the boundary is the
conformal measure of the boundary.
 Darcy's law (\ref{1})
   reads: {\it the complex velocity  of the boundary
is proportional to the conformal measure}. In other words,
  the complex potential of the flow is a conformal map of the outer
domain to the cylinder. Equivalently, the Cauchy-Riemann conjugate
of  pressure, $\xi(x,y)$, is the {\it stream function}.

\begin{figure}[h!!!!!!] \begin{center}
 \includegraphics*[width=6cm]     {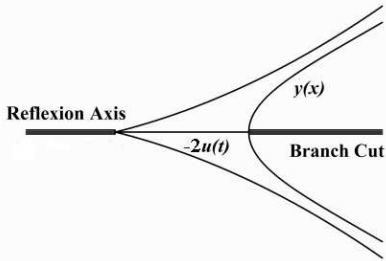}
\caption{\label{cusp}
A finger extends to the right of  its boundary
$y(x)$.\,$-2u(t)$ is the distance from the tip to the singular point.
}
\end{center} \end{figure}

For simplicity, let the bubble
and finger have a symmetry axis.
In Cartesian coordinates, the boundary
can be  described by a
multivalued function $y(x)$, Fig~\ref{cusp}.

Another useful way to describe the boundary is through 
the height function $h(z)$. It  is 
an analytic function in some vicinity of the boundary,
 taking values $h(x+iy)=y(x)$ on the boundary .
With the help of  the Schwarz reflection principle,
  we obtain $h(\phi)-\overline{ h(-\bar\phi)}=-i(z(\phi)-
\overline{  z(-\bar\phi))}$, where 
$z(\phi)$ is the inverse of the map 
(we write $h(\phi )=
h(z(\phi ))$).
Therefore,
\begin{equation}\label{41}
2ih(\phi)= z(\phi)-\overline{  z(-\bar\phi)}.
\end{equation}
We will also use a generating  function $\partial_z \omega(z)=-2i h(z)$.
Darcy's law can be interpreted as evolution of the height function.
The complex velocity  
$v=\dot{\bar z}$
can be analytically  continued into the outer domain, as
$v=\p_t \p_z\omega(z)=-2i\p_t h(z)$.
Then from $v = -2i\p_z \phi$, Darcy's law becomes
\begin{equation}\label{44a}
\p_t h(z)=\p_z\phi(z).
\end{equation}

\paragraph*{3. Branch points}
If the boundary is analytic,
the map  $\phi(z)$ and   the height function $h(z)$ can be
analytically continued inside the bubble,
 until they reach  singularities. For an important class
of analytic boundaries, 
the conformal map $z(\phi )$ is a rational function of
$e^{i\phi }$.
Such domains are called algebraic.
In this case, the singularities of the inverse map
are generically simple branch points
$z_0(t),\dots,z_{q-1}(t)$
(analogs of Riemann invariants in hydrodynamics),
  where  $\p_z\phi\sim (z-z_i(t))^{-1/2}$.
Hele-Shaw flows preserve the class of algebraic domains \cite{Tavveer}.

As the boundary moves, so do the branch points. We identify the traces of the branch points 
with the branch cuts. Comparing
  singularities  of  (\ref{44a}), we get a system of
coupled equations for the branch points,
\begin{equation}\label{4}
\dot z_i=\frac{\p\phi}{\p h}\Big|_{z_i},\quad i=0,1,\dots,q-1.
\end{equation}
\paragraph*{4. Finite time singularities}
In full generality, these equations are difficult to analyze.
Qualitatively,  branch points tend to move towards
 the boundary. As a result, the  boundary is  pushed away,
forming a finger. When, eventually,
a  branch point reaches the boundary, the finger forms a
cusp-like singularity. It appears that
for almost all initial algebraic domains, a cusp-like singularity
 emerges at a finite
time,  i.e. at a finite area of the bubble.
The idealized law (\ref{1}) then needs to be 
corrected by regularizing the singularity \cite{Mark}.
Some examples of singular Hele-Shaw flows were elaborated
in \cite{sing}.

In this letter we suggest a unified approach to the finite time
singularities based on a singular limit of the KdV
integrable hierarchy.
We concentrate on the critical regime of the flow,
when fingers are close to a cusp. Generically,
an isolated branch point, say $z_0=(x_0,0)$,  is found very close to
  the boundary (tip of the finger),
  i.e., $\phi(z_0)\approx 0$.
From (\ref{4}), this branch point (and tip) moves 
with velocity $\dot x_0={\partial\phi}/{\partial y}|_{x_0}$.

The origin of the finite time singularities can already be seen
  from this equation by means of scaling analysis.
  Around the tip, the map $z(\phi)$ has a regular expansion
$z(\phi)-x_0= ia\phi+b\phi^2+ic\phi^3+\dots$, with  real
$a,b,c \ldots \, $.
It follows from the reflection formula (\ref{41}) that  $h(\phi)=
a\phi+c\phi^3+\dots$.
  If a branch point is close to the tip, then $a$ is small,
while  $b$ and $c$ are of order 1.  If there is only one scale,
$u(t)$,  then $x_0$, measured from  what will be the cusp tip,
and $a$ are of the same order $u(t)$. The velocity
of the finger tip $\dot x_0\sim \dot u$ grows
with the curvature  $a^{-1}\sim u^{-1}$ of the tip such that
  $\dot u\sim u^{-1}$, hence  the scale vanishes as  $u\sim (t_c-t)^{1/2}$.
  At the critical  time $t=t_c$, $u=0$  and the curve forms a
 (2,3)-cusp, $y(x)\sim x^{3/2}$.

More general cusps are characterized by two integers
$(q,p),\,q<p$, implying that the finger
  is bounded by
$|y(x)| \sim x^{p/q}$, $x>0,\; q\;\mbox{is even}$.
Higher cusps correspond to
higher order singularities merging at the boundary.

In what follows we mostly consider the case $q=2$,
making only brief remarks on the higher cusps.
The details will be given elsewhere.

\paragraph*{5. Critical regime and scaling functions.}
When the finger  is very close to becoming a cusp, the rest of the
  bubble  does not affect its evolution. We will call this
regime {\it critical}.  In the critical regime, different
  scales separate.  It is convenient to define the scales in
the complex plane of $\phi$.
At  $\phi = O(u^{1/2})$ we see the details of the  tip. At
$1 \gg \phi \gg u^{1/2}$, the details of the
tip cannot be seen, the finger looks like a cusp.
Finally,  $\phi = O(1)$ corresponds to the rest of the body
of the bubble.  The scale $u(t)$ changes with time and eventually
disappears at the critical point. In that limit, the finger is scale invariant.

To summarize, in the critical regime $u \to 0, \, t \to t_c$,
  the inverse map and the  height function
  are
\begin{equation}\label{31}
       z(\phi)\! -\! ih(\phi)=u^{\frac{q}{2}}Q({\phi}/{u^{1/2}}),\;
h(\phi)=u^{\frac{p}{2}}P(\phi/u^{1/2})
  \end{equation}
  where the scaling functions $Q$ and $P$ do not depend on $u$.
At $q=2$, $Q$ is an even quadratic polynomial and the
scaling ansatz holds
everywhere in the critical region.

The asymptotic behavior of the  finger, $y\sim \pm x^{p/q}$,
allows us to
identify the scaling functions.
They are polynomials of degrees $q$ and $p$,
respectively.
The reflection symmetry suggests that $Q$ is an even polynomial,
while $P$ is
odd, and that all their coefficients are real.
This follows by matching asymptotes
in different regions: far from the tip, the finger asymptote is
$y\sim \pm x^{p/q}$. Since $y \ll x \ll 1$, we approximate $y\sim z^{p/q}$,
where $z=x+iy$. Higher order corrections will make $y$ real. Therefore, we
conclude that the height function, in the domain close to the boundary,
but away from
the tip, is $h(z)\sim z^{p/q}$.  This also means that  at
$u^{1/2} \ll \phi \ll 1$, and  close to the real axis,
$Q(\phi )\sim \phi^q,\,P(\phi)\sim \phi^p$.
Close to the tip, where $\phi \ll u^{1/2}$,
both $z(\phi)$ and $h(\phi)$ are
regular in $\phi$.
The only holomorphic functions  matching  these conditions
are polynomials.


\paragraph*{6. }
An important remark on the scaling ansatz is in order.
From (\ref{31}), approximating the univalent  map $z(\phi)$, being
by polynomials makes it no longer  univalent at $q>2$.
It  covers the $z$-plane  $q$ times, as many as
the number of sheets of the complex curve.
This probably means that  at $q>2$ the scaling ansatz does not hold
everywhere in the upper half-plane of $\phi$.
It holds only in certain sectors of the plane adjacent to the
real axis  and breaks down otherwise.
Their union, covering roughly
$2/q$ part of the upper half-plane - {\it {
a physical domain}}
determines a
physical branch of the map $\phi (z)$. 
The physical branch  must obey the
reflection symmetry and be real
at the boundary of the finger.
This is impossible unless the stream
function $Re \phi$ is allowed to have
a discontinuity across the reflection axis outside the finger.
Every monomial $z^{n/q}$ in the asymptotic expansion
$\phi(z)=z^{1/q}+\mbox{negative powers of $z^{1/q}$}$
should be understood as
$ z^{1/q}$, if $0<\mbox{Arg} z<\pi-\epsilon$, and
$ e^{2\pi i/q}z^{1/q}$, if $-\pi+\epsilon<\mbox{Arg} z <0$. 
As a result
the
imaginary part of  $\phi(z)$ has a finite discontinuity  
on the  reflection line, at least far
away from the tip, unless $q=2$.  

This  indicates that cusps with $q>2$ do not appear for 
algebraic  simply-connected domains.  We do not have an 
interpretation of this phenomenon. 
The non-univalidness of the map
suggests that there is another finger 
at the left, while the cusp singularity
implies their merging.
 


\paragraph* {7. dKdV hierarchy}
From  (\ref{31}), in the scaling limit,
the  generating function has the asymptotic expansion
$\omega(z)-\omega(z_0)=-2i\int ^z_{z_0} h(z') dz' + O(u^p)
\approx u^\frac{p+q}{2}\int_0^{\phi} PdQ$. It
is a  truncated Laurent series in $z^{1/q}$,
\begin{equation}\label{44}
\omega(z)=i\sum_{n=1}^{p+q} t_n  z^{n/q}+\mbox{negative powers of
  $z^{1/q}$}.
\end{equation}
We will see in a moment that  all the coefficients $t_k$ except
  $t_1$ are conserved.  Moreover, the coefficient $t_1$ is
proportional  to time,  measured  from the moment of singularity.
  The coefficients $t_k, t_1$  are called 
{\it deformation} and $evolution$ parameters, respectively.
The generating function  and the height function
can be expressed through deformation parameters. As
   polynomials of $\phi$, they are
\begin{equation}\label{y1}
\omega=i\sum_{n=1}^{p+q}t_n\,
\omega_n,
\quad
h=\sum_{n=1+q}^{p+q}\frac{n}{q} t_n\,
\omega_{n-q}.
\end{equation}
Here $\omega_n(\phi)$ is the polynomial
part of $u^{n/2}Q^{n/q}(\phi )$.

Now we can describe the evolution of the curve with respect to all the
deformation parameters. The arguments are borrowed from
\cite{Takasake} with  little changes.
We  note
that $\p_{t_n} \omega$ is a polynomial of
degree $n$ in $\phi$.  As a Laurent series
in  $z^{1/q}$, it   has only  one
term, $z^{n/q}$, with positive degree.
The only polynomial of this kind is $\omega_n$. Therefore,
\begin{equation}\label{xy}
 \p_{t_n}
 h(z)=\p_z \omega_n(z).
\end{equation}
Notice that the flow in real time (\ref{44a}) appears on the same
footing as flows with respect to the deformation parameters. Setting
$n=1$, we recover the flow equation (\ref{44a}).

Eqs.  (\ref{44a},\ref{xy})  can be cast in the form of flow equations,
if one takes time derivative
at constant $\phi$.  Defining the Poisson bracket
with respect to the canonical pair $t, \, \phi$
such that $\{f,\,g\}=\p_t f\p_\phi g-\p_t g\p_\phi f$,
eqs. (\ref{44a},\ref{xy}) read
$
\{h, \, z\} = 1, \,\, \p_{t_n}
  h(\phi)= \{ h, \, \omega_n  \}.
$
From this we conclude that the inverse map evolves 
as
\begin{equation} \label{kdv}
\p_{t_n} z = \{z , \, \omega_n \}.
\end{equation}
  Compatibility of these equations gives a
closed set of nonlinear equations for the
coefficients of the polynomials (\ref{31})
$\p_{t_m}\omega_n -\p_{t_n}\omega_m =\{\omega_n,\,\omega_m\}.$
At $q=2$, it is the dispersionless
KdV hierarchy (dKdV). At $q>2$, equations (\ref{kdv})
constitute the dispersionless Gelfand-Dikii hierarchy \cite{Takasake}. 
\paragraph*{8. Solutions of the dKdV hierarchy}
At $q=2$, a complete solution is available.
 In this case $z(\phi)=\phi^2-2u$, the evolution
does not depend on
$t_{2n}$ and
\begin{equation} \label{odd}
\omega_{2n+1} =
\sum_{k=0}^{n}\frac{(2n+1)!!}{(2n-2k+1)!!}
\frac{(-u)^k}{k!}\phi^{2n-2k+1}.
\end{equation}
Equations (\ref{kdv}) become
$$\p _{ t_{2n+1}} u= \frac{(2n+1)!!}{n!}(-u)^n \p _{ t_1}u.
$$
The first equation is the
familiar Hopf-Burgers equation
\begin{equation} \label{15}
\p_{t_3} u + 3u\p_{t_1}u=0.
\end{equation}
The hodograph
transformation gives a general solution 
\begin{equation} \label{hodograph}
 \sum_{k=0}^{l}\frac{(2k+1)!!}{k!}t_{2k+1}(-u)^k = 0,\quad p=2l+1.
\end{equation}
More details of the application
to the singular limit of the Hele-Shaw flow
can be found in \cite{25}. Some results
for $q>2$ are discussed in \cite{scaling}.

\paragraph*{9. KdV hierarchy}
  Eq. (\ref{15}) is a singular $\hbar\to 0$ limit of the
full dispersive KdV equation
\begin{equation} \label{18}
4\p_{t_3} u+12u\p_t u+\hbar^{5/2}u_{ttt} =0.
\end{equation}
To clarify the
nature of this limit, we recall the definition of the
$q$-reduced KP hierarchy. It
is a set of nonlinear  equations compactly written through a   pair
of operators
$L, M$, differential  operators
in time of degrees $q$
and $p$:
\begin{equation}
L=\hat\phi^q-\sum_{l=1}^q [
e_l,\hat\phi^{q-l}]_{+},\quad M=\sum_{n=1+q}^{p+q}\frac{n}{q} t_n
\Omega_{n-q}.
\end{equation}
Here $[\, , \, ]_{+}$
denotes the anti-commutator,
$\hat \phi =\hbar\p_t$, and the coefficients  $e_l$ are
functions of a string of  ``times" $t_1,\dots,t_{p+q}$, and
  $\Omega_n=L^{n/q}_+$ is a differential operator of degree $n$,
obtained from the pseudo-differential operator $L^{n/q}$ by omitting a
non-differential part. The dependence of the coefficients $e_l$ on the ``times" is introduced by the
Lax-Sato equations
$\hbar\p_{t_k}L=[L, \,\Omega_n]$,
$\hbar\p_{t_k}M=[M, \,\Omega_n]$,
supplemented by the condition
$[L,\,M]=\hbar$.
An alternative way to define the hierarchy
is to impose the consistency conditions
$\p_{t_n}\Omega_m-\p_{t_m}\Omega_n=[\Omega_m,\,\Omega_n]$.

The dispersionless (``quasiclassical") limit of this
hierarchy is obtained by
 replacing the differential operator $\hat\phi=\hbar\p_t$ by a
function
$\hat\phi\to\phi$, and the commutator $i\hbar^{-1}[\,,\,]$ by the Poisson
bracket $\{\,,\,\}$. We have seen that
in this limit the formal objects of the
Lax-Sato construction acquire a
clear physical interpretation in terms of the
Hele-Shaw flow. Namely,
$\hat\phi, L$ and $M$  become respectively the complex potential $\phi$,
the coordinate
$z$, and the height function $h$. This correspondence also gives a
meaning of $\hbar$ as a quantum of area.

\paragraph*{10. Applications} The integrable structure
revealed in this letter allows one to use the KdV theory in the study of
the Hele-Shaw and
other moving boundary problems.
Special solutions of the
hierarchy describe (i) bubble coalescence,
(ii) bubble break-off, (iii) branching,
  (iv) bubble creations, etc. We report some of them in \cite{25}.


It is known that the dispersionless limit of
non-linear waves is singular \cite{singular}. The solutions
suffer from
nonphysical shock waves. Smooth initial data
generally evolve into a multi-valued
(``overhanged") function within a finite time.
Such nonphysical solutions are
equivalent to the finite-time singularities of the
Hele-Shaw flow. 

The singular behaviour should be
corrected  by a regularization of the cusps at short
distances, by introducing a new scale -- a short
distance cut-off. Surface tension, lattice or DLA-like
regularizations were
considered. 


The integrable structure of  singularities 
suggests a novel, ``dispersive" regularization. 
The  flow (\ref{1}) is seen as an ill limit of the   true
dispersive flow, just as Eq. (\ref{15}) is seen as a singular
limit of   Eq. (\ref{18}). We will present details of this
regularization elsewhere. Here we only  note that this regularization
treats the Hele-Shaw flow as a stochastic process of
deposition of small
particles with an  area
$\hbar$. It carries a similar physics as the DLA \cite{DLA}. 
\paragraph*{Acknowledgments}
We were  benefited by discussions
 with  I. Krichever, A. Marshakov
  and M. Mineev-Weinstein. P. W. and A. Z. are particularly grateful
 to E. Bettelheim and O. Agam
   for numerous discussions, help, and contribution to our understanding of
  physics of Hele-Shaw
 flow, and I. Loutsenko for a critical comment. R.T. thanks I. Aleiner and A. Millis for support. P.W. and R.T. 
were supported by the NSF MRSEC Program under
DMR-0213745, NSF DMR-0220198.  A.Z. was also supported in  part by  RFBR grant
03-02-17373, by grant INTAS 03-51-6346 and by the
grant 
NSh-1999.2003.2. 


\end{document}